\documentclass[twocolumn]{aastex631}

\usepackage{graphicx}	
\usepackage{amsmath}	
\usepackage{amssymb}	
\usepackage{xcolor}
\usepackage{cancel}
\usepackage{multirow}
\usepackage{bigdelim}

\newcommand{\mcl}{M_{\rm cl}}
\newcommand{\msolt}{M$_\odot$~}
\newcommand{\mpcsqt}{M$_\odot$~pc$^{-2}$}

\newcommand{\be}{\begin{equation}}
\newcommand{\ee}{\end{equation}}
\newcommand{\ie}{\emph{i.e.}, }
\newcommand{\eg}{\emph{e.g.}, }

\definecolor{color1}{HTML}{FF6C0C}

\newcommand{\rev}[1]{\textcolor{black}{#1}}

\received{2021 September 23}
\revised{2021 December 21}
\accepted{2021 December 30}

\submitjournal{ApJ Letters}

\shorttitle{Bursting Superbubbles}
\shortauthors{M. E. Orr et al.}

\usepackage{totcount}
\newtotcounter{citenum}
\def\oldcite{}
\let\oldcite=\bibcite
\def\bibcite{\stepcounter{citenum}\oldcite}

\begin{document}


\title{Bursting Bubbles: Clustered Supernova Feedback in Local and High-redshift Galaxies}

\correspondingauthor{Matthew Orr}
\email{matt.orr@rutgers.edu}

\author[0000-0003-1053-3081]{Matthew E. Orr}
\affiliation{Department of Physics and Astronomy, Rutgers University, 136 Frelinghuysen Road, Piscataway, NJ 08854, USA}
\affiliation{Center for Computational Astrophysics, Flatiron Institute, 162 Fifth Avenue, New York, NY 10010, USA}
\affiliation{TAPIR, Mailcode 350-17, California Institute of Technology, Pasadena, CA 91125, USA}

\author[0000-0003-3806-8548]{Drummond B. Fielding}
\affiliation{Center for Computational Astrophysics, Flatiron Institute, 162 Fifth Avenue, New York, NY 10010, USA}

\author[0000-0003-4073-3236]{Christopher C. Hayward}
\affiliation{Center for Computational Astrophysics, Flatiron Institute, 162 Fifth Avenue, New York, NY 10010, USA}

\author[0000-0001-5817-5944]{Blakesley Burkhart}
\affiliation{Department of Physics and Astronomy, Rutgers University, 136 Frelinghuysen Road, Piscataway, NJ 08854, USA}
\affiliation{Center for Computational Astrophysics, Flatiron Institute, 162 Fifth Avenue, New York, NY 10010, USA}

\begin{abstract}
We compare an analytic model for the evolution of supernova-driven superbubbles with observations of local and high-redshift galaxies, and the properties of \rev{intact} HI shells in local star-forming galaxies.
Our model correctly predicts the presence of superwinds in local star-forming galaxies (e.g., NGC 253) and the ubiquity of outflows near $z \sim 2$.  We find that high-redshift galaxies may `capture' 20-50\% of their feedback momentum in the dense ISM (with the remainder escaping into the nearby CGM), whereas local galaxies may contain $\lesssim$10\% of their feedback momentum from the central starburst.  Using azimuthally averaged galaxy properties, we predict that most superbubbles stall and fragment \emph{within} the ISM, and that this occurs at, or near, the gas scale height.  
\rev{We find a consistent interpretation in the observed HI bubble radii and velocities, and predict that most will fragment within the ISM, and that those able to break-out originate from short dynamical time regions (where the dynamical time is shorter than feedback timescales).}
 Additionally, we demonstrate that models with constant star cluster formation efficiency per Toomre mass are inconsistent with the occurrence of outflows from high-$z$ starbursts and local circumnuclear regions. 
\end{abstract}



\keywords{Supernova remnants (1667) --- Superbubbles (1656) --- Star formation (1569) --- ISM (847) --- Galaxy evolution (594) --- Stellar Feedback (1602)} 



\section{Introduction}

Turbulence driven by supernova (SN) explosions provides \rev{a critical source of} support in galaxies that \rev{helps set} the disk scale height and prevents gravity from causing runaway star formation \citep{Faucher-Giguere2013}\rev{, alongside other physical processes like gas accretion onto and transport within galaxies\footnote{\rev{Transport processes are especially important at high redshifts $z~\gtrsim~1$, when galaxies are most rapidly accreting gas and assembling their stellar populations \citep{Behroozi2013b,Krumholz2018}.}} \citep{Krumholz2018}}.  Other forms of feedback, from stellar winds to photoionizing radiation, are important in the context of setting the \emph{local} efficiency of star formation in molecular clouds \citep{Dale2014, Grudic2018, Li2019,Burkhart2019}, but are unable to affect structures on the scale of galactic disks.  Therefore, in order to make sense of the turbulent structure of the ISM on the largest scales and how feedback processes drive galactic winds and fountains, a rigorous understanding of the nature and effects of supernova feedback is required.

Star formation is inherently a clustered process, with stars forming hierarchically inside marginally gravitationally bound molecular clouds \citep{Lada2003}.  Star-forming clouds are also clustered temporally, only producing stars for roughly an internal free-fall time, ($\sim$1-3 Myr; the clouds themselves existing for a few free-fall times, $\lesssim 10$ Myr), before being dispersed initially by `prompt' feedback processes, like stellar winds and photoionizing radiation, and finally by the first core-collapse supernova explosions \citep{Murray2010, Grudic2018, Li2019}.  As stellar explosions continue in the molecular environment the overlapping supernova remnants can form an encompassing shock-front described as a superbubble \citep{Koo1992}.

Galaxy simulations with realistic implementations of star formation and supernovae are able to reproduce observed stellar mass relations, low average star formation efficiencies, and the level of turbulence in galaxies \citep[][]{Wetzel2016, Agertz2016, Kim2017, Pillepich2018, Hopkins2018:fire, Orr2018, Orr2020}.  These simulations implicitly incorporate clustered supernova feedback as the star formation events themselves are inherently clustered.  A number of `small-box' simulations have focused on the ability of supernova-driven superbubbles to drive galactic fountains and outflows \citep{Martizzi2016, Kim2017, Fielding2018}, confirming that the clustering of supernovae is crucial to realizing realistic ISM structure.  However, until now there has not been a robust, first principles model of \emph{how} clustered supernovae in galaxies regulate star formation and drive outflows, and specifically the local interstellar medium (ISM) conditions required for either. 

In \rev{\citet{Orr2021b:arx}}, Paper I of this series, we developed an analytic model of clustered supernova feedback, and established the likely outcome and effects of supernova-driven superbubbles in the ISM, and on the flow of gas into and out of galaxies.  We found that the local gas fraction and dynamical time in galaxies alone determine if star clusters can drive galactic winds/outflows (\rev{the subsequent wind/outflow properties, including the ability to entrain cold material, being described by a wind-specific model like that of} \citealt{Fielding2021}), with implications for the stall/fragmentation scale of superbubbles, and the effective strength of feedback in driving gas turbulence within galaxies.

In this letter, we will compare our model from Paper I (hereafter, O21) of (spatially and temporally) clustered core-collapse SNe feedback, in the form of superbubbles expanding into a galactic gas disk, with observations of galactic outflows/fountains and HI holes. In \S~\ref{sec:model}, we briefly reiterate our simple model for superbubbles, and the possible outcomes of the evolution of those bubbles in the ISM of galaxies.  We then compare our model directly with observations and simulations in \S~\ref{sec:observations}. Finally, we discuss the model in the context of star formation/galaxy evolution literature, and summarize our results in \S~\ref{sec:discussion}.  
\section{Superbubble Model, In Brief}\label{sec:model}

We review here our model (the details of which can be found in O21) for the growth and eventual outcomes of a supernova-driven superbubble following the formation of a star cluster of mass $\mcl$ in a GMC.  A short period after the formation of the star cluster, core-collapse SNe (hereafter referred to simply as SNe) begin to occur as the most massive stars end their lives.  For our model, we assume the star cluster to form instantaneously, with a formation efficiency scaling with the local gas surface density (per the simulations of \citealt{Grudic2018}).  Approximately $N_{\rm SNe} \approx \mcl/100$ M$_\odot$ supernovae detonate over a short period of $t_{\rm SNe} {\sim}40$ Myr, with the supernovae remnants overlapping to form a cavity in the ISM that expands as a superbubble.  The superbubble expands until the shock-front either comes into pressure equilibrium with the surrounding ISM, or breaks out of the gas disk and drives a galactic fountain/outflow (see \S~\ref{sec:outcomes}).  The cartoon in Figure~1 in O21 illustrates the  model and general outcomes of the superbubble evolution.

\subsection{Superbubble Evolution in the ISM}\label{sec:remnantevolution}


\begin{table*}\caption{Superbubble Outcome Boundaries}\label{table:boundaries}
\centering
\begin{tabular}{llll}
\hline
Boundary Cases (Description) & Boundary Equation & Parametric Constraint & O21 Eq.~No.\\
 \hline  \vspace{-3mm}
&&\\ \vspace{-2mm}
\textbf{PBO/PS Case}	& $\tilde f_g = \frac{\sqrt{2}\pi G}{3} \frac{\Sigma_{\rm crit}}{(P/m_\star)_0} \left(\frac{4 \Omega t_{\rm SNe}}{2-\alpha}\right)^{1-\alpha} \frac{1}{\Omega}$ & $\Omega \geq (2 -\alpha)/4t_{\rm SNe}$ & Eq.~9 \\ 
 (``Powered Break-out/Stall'') &&\\ \vspace{-2mm}
\textbf{PBO/CBO Case} 		& $\tilde f_g = \frac{\sqrt{2}\pi G}{3} \frac{\Sigma_{\rm crit}}{(P/m_\star)_0} \left(\frac{2-\alpha}{4 \Omega t_{\rm SNe}}\right) \frac{1}{\Omega}$ &\rdelim\}{5}{0pt}[~$\Omega \leq (2 -\alpha)/4t_{\rm SNe}$] & Eq.~12 \\
(``Powered/Coasting Break-out'')&&\\ \vspace{-2mm}
\textbf{CBO/CF Case}		& $\tilde f_g = \frac{\sqrt{2}\pi G}{3} \frac{\Sigma_{\rm crit}}{(P/m_\star)_0} \frac{1}{\Omega}$ & & Eq.~15 \\ 
 (``Coasting Break-out/Fragmentation'')&&\\ \vspace{-2mm}
\textbf{CF/PS Case}	& $\tilde f_g = \frac{\sqrt{2}\pi G}{3} \frac{\Sigma_{\rm crit}}{(P/m_\star)_0} \left(\frac{4 \Omega t_{\rm SNe}}{2-\alpha}\right)^3 \frac{1}{\Omega}$ & & Eq.~13\\ \vspace{2mm}
 (``Coasting Fragmentation/Powered Stall'') &&\\ 
 \hline 
\end{tabular}
\end{table*}

In O21, we considered a simplified slab geometry for the ISM with a mean mass-density of $\bar \rho_g = \Sigma_g/2H$, where a star cluster forms at the galactic mid-plane with a mass $M_{\rm cl} = \pi H^2 \Sigma_g^2/\Sigma_{\rm crit}$ (according to \citealt{Grudic2018}, $\Sigma_{\rm crit} = 2800$ \mpcsqt).  To model the evolution of the superbubble, we take the bubble to be in a momentum-conserving phase, with the momentum of the shock-front at a radius $R_b$, having swept up the mass of gas within that radius, to be $P_b = \frac{4}{3}\pi R_b^3 \bar\rho_g {\rm d}R_b/{\rm d}t$ (Eq.~3, O21, also \citealt{Fielding2018, El-Badry2019}), where ${\rm d}R_b/{\rm d}t \equiv v_b$ is the expansion velocity of the superbubble.  At all points in time, we balance this momentum with the cumulative momentum injected up until that time by SNe from the central star cluster (\ie $P_b = P_{\rm SNe}$).  To model the rate and nominal momentum injection of supernovae, we invoke a power law delay time distribution, with d$N_{\rm SN}/$d$t \propto t^{-\alpha}$ (see Appendix A of \citealt{Orr2019} and O21 for a more detailed discussion; in this Letter we take $\alpha = 0.46$), and assume that a fiducial momentum of $(P/m_\star)_0$ is injected by each supernova ($\approx 3000$ km/s, \citealt{Martizzi2015}, normalized as 1 SN per 100 M$_\odot$).  Notionally, all SNe occur over a time period $0 < t < t_{\rm SNe}$ corresponding to the time from first SNe to occur in the star cluster (lifetime of the most massive star formed) until the time of the last SNe to occur (lifetime of the least massive star to undergo a core-collapse SN, $\approx 40$ Myr), which divides up our parameter space into cases where the bubble evolves with and without additional supernova momentum being injected.

Balancing the bubble and feedback momenta, we found relations bubble radius and shock-front velocity in time (Eqs.~7 \& 8, O21):
 \[
  \frac{R_b}{H} = \left[ 6 \frac{\Sigma_g}{\Sigma_{\rm crit}} \frac{(P/m_\star)_0}{H/t_{\rm SNe}}  \right]^{\frac{1}{4}}
  \begin{cases}
  
  \!\begin{aligned}
       &\frac{1}{(2-\alpha)^{1/4}} \left( \frac{t}{t_{\rm SNe}}\right)^{\frac{2-\alpha}{4}}\\
       & \qquad \qquad 0 < t < t_{\rm SNe}
    \end{aligned} & \\
    \!\begin{aligned}
       &\left[\frac{t}{t_{\rm SNe}} -\left(\frac{1-\alpha}{2-\alpha}\right) \right]^{\frac{1}{4}}\\
       & \qquad \qquad t > t_{\rm SNe}
    \end{aligned}           
    
  \end{cases}
\]

 \[
  \frac{v_b}{\sigma} =\frac{1}{4\Omega t_{\rm SNe}}\left[ 6 \frac{\Sigma_g}{\Sigma_{\rm crit}} \frac{(P/m_\star)_0}{H/t_{\rm SNe}}  \right]^{\frac{1}{4}}
  \begin{cases}
  
  \!\begin{aligned}
       &(2-\alpha)^{\frac{3}{4}} \left( \frac{t}{t_{\rm SNe}}\right)^{-\frac{(2+\alpha)}{4}}\\
       & \qquad \qquad 0 < t < t_{\rm SNe}
    \end{aligned} & \\
    \!\begin{aligned}
       & \left[\frac{t}{t_{\rm SNe}} -\left(\frac{1-\alpha}{2-\alpha}\right) \right]^{-\frac{3}{4}}\\
       & \qquad \qquad t > t_{\rm SNe}
    \end{aligned}           
    
  \end{cases}
\]


\subsection{Superbubble Outcomes}\label{sec:outcomes}
\begin{figure*}
\centering
	\includegraphics[width=0.8\textwidth]{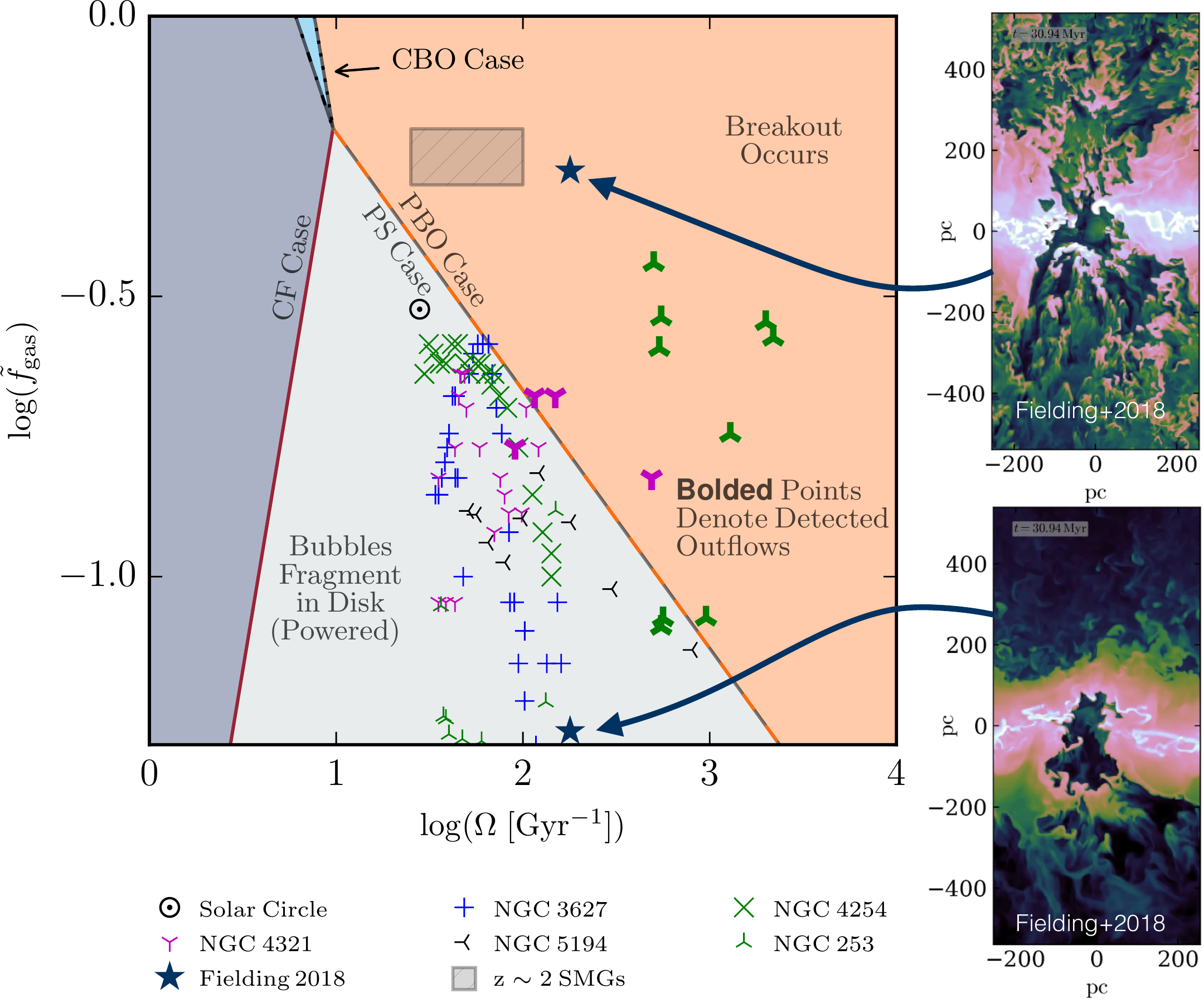}
    \caption{Gas fraction--dynamical time phase space of remnant outcomes, following Eqs.~9, 12, 13 \& 15 from O21.  Low-redshift observations compiled from \citet[][Solar Circle estimate]{McKee2015} \citet[][radial gas fractions for NGCs]{Gallagher2018}, \rev{\citet[][for NGC 253 circumnuclear clump surface densities]{Leroy2015a}, \citet[][for molecular gas data in the NGC 253 disk]{Sorai2000}}, and rotation curves from \citet{Sofue1999}, \citet{Chemin2006} \& \citet{Dicaire2008}. Intermediate-redshift SMG observational estimates inferred from \citet{Tacconi2013} and \citet{Genzel2020}. Bolded NGC points denote detected outflows.  Solid orange and light-blue regions denote cases where remnant successfully breaks out of the disk, and light-grey and blue-grey regions correspond to cases where the remnant stalls/fragments in the ISM.   We also compare with superbubble simulations from \citet[][]{Fielding2018}, which overlap in their initial disk and star cluster properties with this model.  \emph{Insets at right}: These simulations are in good agreement with our model, hosting dramatic break-outs in \textbf{PBO case} parameter space and churning, stalled bubbles in \textbf{PS case} parameter space (images are of projected gas density for the simulations from \citealt{Fielding2018}, adapted from their paper).}
    \label{fig:case}
\end{figure*}

The outcomes of the evolution of these superbubbles are broken down into four cases, depending on whether or not the bubbles break out of the disk (at a time $t_{\rm BO}$), and if the central star cluster is still producing SNe.  If the superbubble comes into pressure equilibrium with the surrounding ISM (for a turbulent ISM, $P_{\rm ISM} \sim \bar \rho_g \sigma^2$), it will not maintain coherence in its expansion to reach the gas disk scale height, and instead the shock-front will fragment/stall (as shown in the simulations by \citealt{Fielding2018}).  And so, we take the superbubbles to stall and fragment when $v_b \approx \sigma$, demarcating the difference between cases where the remnant successfully and unsuccessfully reaches the disk scale height.

We considered the following cases in O21,

\begin{itemize}
\setlength{\itemindent}{-2em}
\setlength{\itemsep}{1.5ex}
  \setlength{\parskip}{0pt}
  \setlength{\parsep}{0pt}   
  
\item[] \textbf{PBO Case:} ``Powered Break-out'', SNe remnant superbubble reaches the gas disk scale height, $R_b = H$, \emph{before} the central star cluster ceases producing SNe, $t_{\rm BO} < t_{\rm SNe}$.

\item[] \textbf{CBO Case:} ``Coasting (unpowered) Break-out'', remnant reaches the gas disk scale height, $R_b = H$, \emph{after} the central star cluster ceases producing SNe, $t_{\rm BO} > t_{\rm SNe}$.

\item[] \textbf{CF Case:} ``Coasting (unpowered) Fragmentation'', the remnant fragments in the turbulent ISM (\ie the velocity of the shock-front falls below the turbulent velocity of the ISM), $v_b < \sigma$, \emph{before} reaching the gas disk scale height, $R_b(v_b = \sigma) < H$, \emph{after} the central star cluster ceases producing SNe, $t(v_b = \sigma) > t_{\rm SNe}$.

\item[] \textbf{PS Case:} ``Powered Stall'', bubble expansion stalls in the turbulent ISM, $v_b < \sigma$, \emph{before} reaching the gas-disk scale-height, $R_b(v_b = \sigma) < H$, \emph{before} the central star cluster ceases producing SNe, $t(v_b = \sigma) < t_{\rm SNe}$.
\end{itemize}

Following our assumptions regarding star cluster formation efficiency, and that locally gas in galaxies finds itself marginally stable against gravitational fragmentation and collapse with Toomre-$\tilde Q_{\rm gas} \approx 1$, we found in O21 that the boundaries between these four cases could be entirely expressed in terms of local gas fraction $\tilde f_g \equiv \Sigma_g/(\Sigma_g+ \Sigma_\star)$ and inverse dynamical time $\Omega \equiv v_c/R$.  We refer to Table~\ref{table:boundaries} for the boundaries between cases, in $\tilde f_g$--$\Omega$ space, along with the constraints in $\Omega$ derived in O21 (and referencing Eq.~numbers therein).

\section{Comparison to Observations}\label{sec:observations}

We compare our model with observational data for the Solar circle, and low-redshift NGCs 253, 3627, 4254, 4321, \& 5194 (M51) in Figure~\ref{fig:case}.  Combining the spatially resolved \rev{molecular gas surface density} data set from \citet{Gallagher2018} \rev{taken with ALMA} for the NGCs (except for NGC 253, which we take \rev{ALMA molecular gas} data from \citealt{Leroy2015a} of the star\rev{-}forming clumps in its circumnuclear region, and \rev{Nobeyama Radio Observatory CO data from \citealt{Sorai2000}} for its disk), from which we derive (azimuthally averaged) radial gas fraction profiles calculated from their data as $\tilde f_{\rm gas} \equiv (\Sigma_{\rm mol} + \Sigma_{\rm HI})/(\Sigma_{\rm mol} + \Sigma_{\rm HI} + \Sigma_\star)$ where $\Sigma_{\rm mol}$ is the molecular gas surface density and $\Sigma_{\rm HI}$ is the atomic hydrogen surface density. \rev{\citet{Gallagher2018} references \citet{Querejeta2015} for radial stellar surface density profiles, whereas we use $J$ and $Ks$ band VISTA data from \citet{Iodice2014} to estimate $\Sigma_\star$ at the location of the star-forming circumnuclear observed by \citet{Leroy2015a}, and stellar disk parameters from \citet{BlandHawthorn1997} for the disk of NGC 253.}  \rev{The \citet{Gallagher2018}} dataset extends \rev{$R = $ 0 -- 6 kpc} radially \rev{(radial bin sizes range from $\sim$180-570 pc in width)}, and so does not extend into the atomic gas-dominated galactic outskirts of any of these galaxies (where dynamical times grow long, see discussion in second half of \S~\ref{sec:critpt}).  \rev{For NGC 253, the disk data from \citet{Sorai2000} has $R=$ 0--5 kpc with $\Delta R \approx $ 200 pc, and the observations of circumnuclear clumps \citep[from][]{Leroy2015a} have $R < 1$ kpc with $\Delta R \sim 30$ pc.} For comparison with our model, we then interpolate the rotation curves for these galaxies measured by \citet{Chemin2006} and \citet{Dicaire2008} to produce inverse orbital dynamical times $\Omega$. Estimates for the Solar circle ($\Sigma_\star \approx 35$ M$_\odot$, $\Sigma_{\rm gas} \approx 15$ M$_\odot$, and $\Omega \approx 35$ Gyr$^{-1}$) are taken from \citet{McKee2015}.

Nearly all of the nearby galaxies (and the Solar Circle) fall into the parameter space of the \textbf{PS case}, with the exception of the central regions ($\lesssim 1$ kpc) of NGCs 253 and 4321, which fall into the \textbf{PBO case} (``Powered Break-out''). None of NGCs 3627, 4254 or 5194 (nor the Solar circle) have significantly detected SN-driven outflows \citep{Calzetti2005, Wezgowiec2012, Law2018}.  \rev{It is difficult to find studies reporting non-detections of outflows (e.g., in NGC 3627 and 5194, we find many reports of gas and star formation rate distributions but no studies of outflow properties), however for NGC 4254 \citet{Wezgowiec2012} reports that a relatively homogenous hot gas distribution (inferred from X-ray emission) disfavors significant star formation-driven outflows.}  However, there \emph{is} evidence of an outflow originating from the circumnuclear region of NGC 4321 \citep[][\rev{supported blueshifted interstellar contamination of NaD absorption in their data}]{Castillo-Morales2007}. And NGC 253 hosts a notable superwind, driven by its central starburst \citep[][\rev{seen as a wide-velocity component molecular CO wind originating from the central starburst}]{Bolatto2013a}. The predicted \textbf{PBO/PS case} boundary is thus consistent with observations in nearby star-forming galaxies with our fiducial parameters.  That the observational data fall nearly along the powered break-out/fragmentation boundary is also consistent with the picture that superbubbles are by and large driving turbulence in galaxies at or near the (gas) disk scale height (see \S~\ref{sec:turbscale} for discussion).

We also include observational estimates for $z\sim 2$ star-forming sub-millimeter galaxies (SMGs), as a hatched grey region.  To compile this data, we combined rotation curve data from \citet{Genzel2020} with (radial) positions of star-forming clumps in a subset of those galaxies from \citet{ForsterSchreiber2011} for a range of $\Omega \sim 25-100$ Gyr$^{-1}$, and then estimates for a range of gas fractions from \citet{Tacconi2013} of $\tilde f_g \sim $ 0.5 -- 0.7.  For nearly any reasonable range of physical parameters, the $z \sim 2$ SMGs appear to fall in the \textbf{PBO case} regime, commensurate with the observed ubiquity of outflows in the intermediate-redshift universe \citep{Weiner2009}.  Furthermore, the transition from galaxies having pervasive dramatic outflows to relatively rarely hosting them may be more a matter of falling local gas fractions than an evolution in $\Omega$.

Lastly, we include two data points from simulations by \citet{Fielding2018}.  These simulations are of a stratified turbulent disk, with $\Sigma_g = 30$ and  $300$ \mpcsqt, an effective disk surface density of $\Sigma_{\rm disk} \approx 570$ \mpcsqt, and an inverse dynamical time of $\Omega \approx 175$ Gyr$^{-1}$.  They span a range of star cluster masses, but we compare with two that fall on the star cluster formation efficiency scaling utilized by our model, namely $M_{\rm cl} = 10^4$ \msolt (when $\Sigma_g = 30$ M$_\odot$ pc$^{-2}$) and $M_{\rm cl} = 10^6$ \msolt (when $\Sigma_g = 300$ M$_\odot$ pc$^{-2}$).  These two simulations fall squarely in \textbf{PS} and \textbf{PBO case} parameter space, respectively, and exhibit the behavior that we expect: the $M_{\rm cl} = 10^4$ \msolt cluster (with its $\Sigma_g = 30$ \mpcsqt~gas surface density) stalls and fragments in the disk, sputtering at times; whereas the $M_{\rm cl} = 10^6$ \msolt cluster ($\Sigma_g = 300$ \mpcsqt) quickly and dramatically breaks out of the disk.  We include images of the projected gas density late in the evolution of these two simulations as insets in Figure~\ref{fig:case}, adapted from their paper.

\subsection{Predicting Turbulence Driving Scale and Effective Strength of Feedback in Observed Galaxies}\label{sec:turbscale}

\begin{figure}
	\includegraphics[width=0.97\columnwidth]{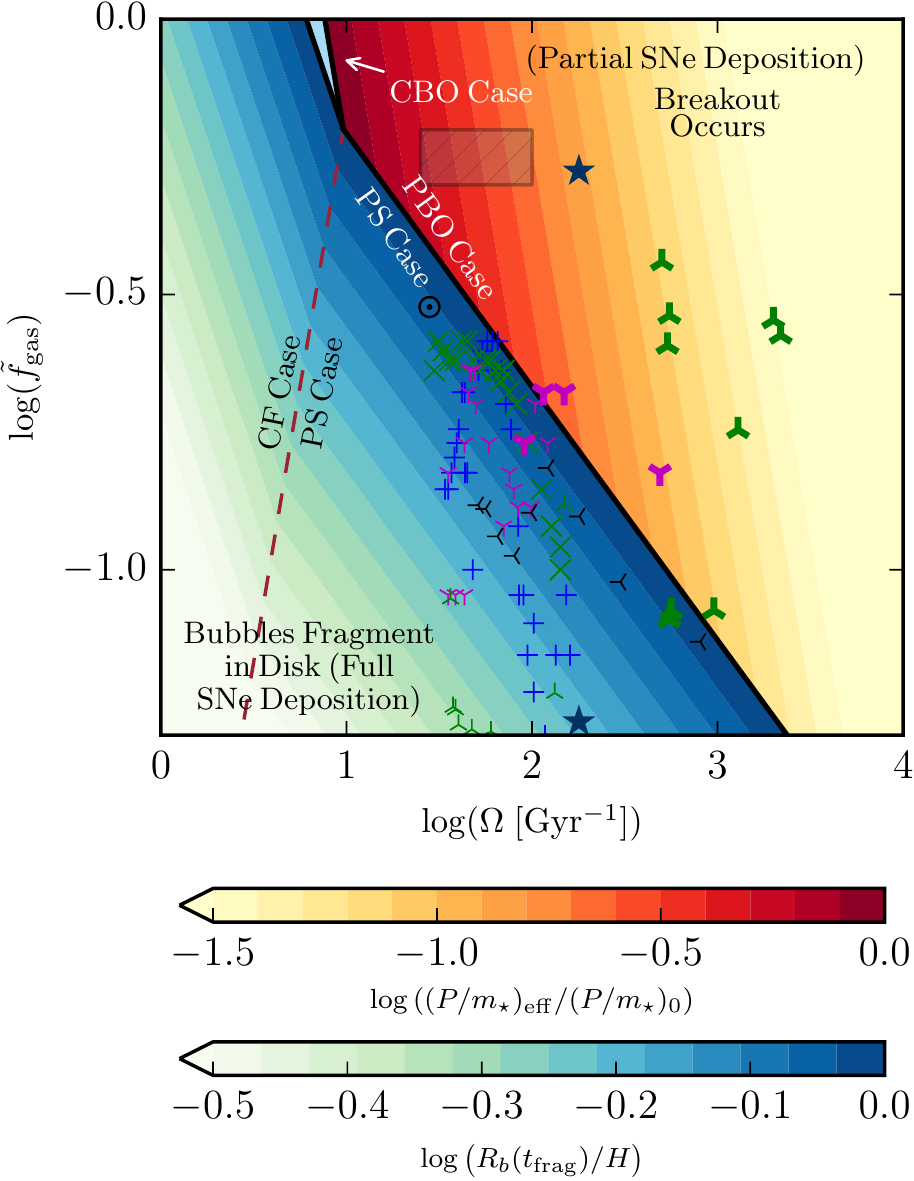}
    \caption{Fragmentation radius as a fraction of disk scale height (for \textbf{CF/PS cases}), and ratio of `effective' to fiducial feedback strength of superbubbles that break out of the disk \emph{before} $t_{\rm SNe}$ (\textbf{PBO case}) in gas fraction--dynamical time phase space. We color the parameter space by predictions from Eqs.~17, 18 \& 20 in O21, and include the same observational data and case boundary lines as in Fig.~\ref{fig:case}.  Light blue region denotes the \textbf{CBO case} where remnant coasts out of the disk but deposits all feedback momentum into the dense gas.  Dashed cardinal line indicates division between the \textbf{CF} (coasting fragmentation) and \textbf{PS} (powered stall) \textbf{cases}.  We find good agreement with the break-out time and fragmentation scale of the simulations by \citet{Fielding2018}.  Low-redshift observations suggest that superbubbles in the local universe fragment very near the gas disk scale height (\ie $R_b(t_{\rm frag})/H \gtrsim 0.7$) and that those which break out deposit as little as $\sim$10\% of feedback momentum \emph{locally} into dense gas.  Whereas, $z \sim 2$ galaxies (hatched grey patch) may deposit the majority of their feedback momentum into the ISM, suggesting that the effects of breakout may be significantly different in high-redshift hosts.}
    \label{fig:fragNpms}
\end{figure}

Here we predict the local turbulence driving scale from the fragmentation of superbubbles within the ISM and \emph{effective} strength of feedback in regions which host bubble break-out, for observed galaxies.  In O21, we calculated the fragmentation/stall radius for bubbles in the \textbf{CF/PS case} and found (Eqs.~17 \& 18, O21),
 
 \[
  \frac{R_b(t_{\rm frag})}{H} = 
  \begin{cases}
  \left[ \frac{(P/m_\star)_0}{\Sigma_{\rm crit}} \frac{3 \Omega \tilde f_g}{\sqrt{2}\pi G} \right]^\frac{1}{3} & {\rm  \textbf{CF Case}} \\
    \!\begin{aligned}
    
       & \left[ \frac{(P/m_\star)_0}{\Sigma_{\rm crit}} \frac{3 \Omega \tilde f_g}{\sqrt{2}\pi G} \right]^\frac{1}{2+\alpha} \\
       & \qquad \times  \left(\frac{2-\alpha}{4\Omega t_{\rm SNe}} \right)^\frac{1-\alpha}{2+\alpha}
    \end{aligned}           & {\rm  \textbf{PS Case}} \\
    
  \end{cases}
\]
Similarly, we predicted the \emph{effective} strength of feedback, $(P/M_\star)_{\rm eff}$, \ie the fraction of momentum deposited into the ISM versus lost to outflows in the event that the superbubble were to break out of the ISM (\textbf{PBO case}), and found (Eq.~20, O21),
\begin{equation*}
\frac{(P/m_\star)_{\rm eff}}{(P/m_\star)_{0}} = \left[ \frac{\sqrt{2}\pi G}{3} \frac{\Sigma_{\rm crit}}{(P/m_\star)_{0}} \frac{2-\alpha}{4 \Omega t_{\rm SNe}} \frac{1}{\tilde f_g\Omega} \right]^{\left(\frac{1-\alpha}{2-\alpha}\right)} \; .
\end{equation*}
This necessarily would affect the slope of the Kennicutt-Schmidt (KS) relation \cite{Kennicutt2012}, in a feedback-regulated framework (\emph{e.g.,} \citealt{Faucher-Giguere2013}), for ISM patches in \textbf{PBO} parameter space.

Figure~\ref{fig:fragNpms} shows the $\tilde f_g$--$\Omega$ parameter space of superbubble outcomes colored by the predicted fragmentation radius and \emph{effective} strength of feedback, for their appropriate cases, with the same observations as Fig.~\ref{fig:case}.   We predict that for local star-forming galaxies (and the conditions of the Solar Circle), most superbubbles which do not break out of the ISM nevertheless still grow to an appreciable fraction of the gas scale height ($R_b(t_{\rm frag})/H \gtrsim$0.7).  Indeed this might be an expected attractor state, as if the gas scale height is to be set by turbulence, and the vertical turbulent field is to be driven by supernovae, then we ought to expect that supernovae have a turbulence driving scale of roughly the scale height.

We see that interestingly, the \emph{effective} strength of feedback is perhaps not dramatically reduced in $z \sim 2$ galaxies, but for local supernova-driven outflows, we predict that \eg NGC 253 might have only $\sim$10\% of the feedback momentum from the central starburst deposited into its ISM.  This suggests that although outflows might be \emph{ubiquitous} at cosmic noon, their effects are significantly different in regards to the ability to locally regulate the ISM.

Comparing to the simulations of \citet{Fielding2018} that fall in the \textbf{PS} \& \textbf{PBO cases}, respectively, we also find satisfactory agreement with our predictions.  Their simulated superbubble that failed to break out grew to $\sim0.9H$, before fragmenting and churning with a relative size of $\sim0.7-0.8H$, and \rev{the simulation that successfully broke out} did so after approximately $\sim$2-3 Myr.  This fragmentation scale was slightly larger than we predict here, but their numerical setup slightly differed from our model assumptions, having a vertically stratified inhomogeneous ISM and flat ($\alpha = 0$) SNe time distribution, which may account for the difference.  In the case of the successful break out simulation, their flat SNe time distribution and $t_{\rm SNe} \approx 30$ Myr may account for the difference between our predicted $(P/m_\star)_{\rm eff}$ and the simulation (see Eq.~19, O21).



\subsection{Comparing to Observed HI Holes \& Bubbles}
\begin{figure}
\centering
	\includegraphics[width=0.95\columnwidth]{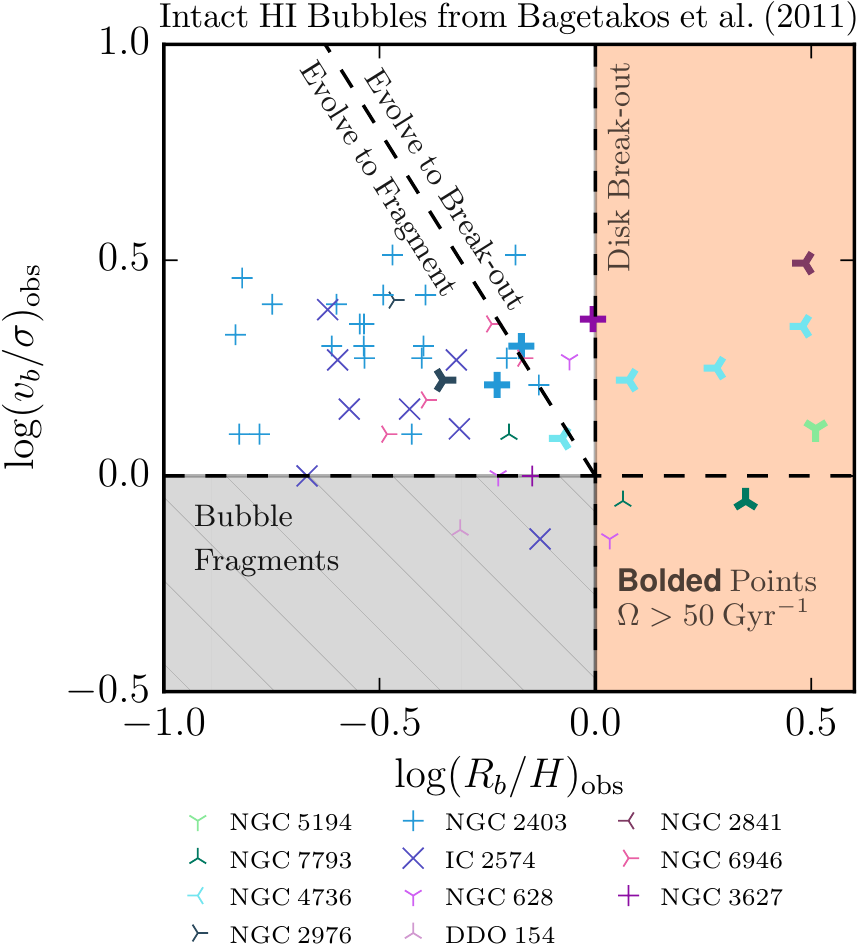}
   \caption{\rev{Observed properties of `intact' HI bubbles from \citet[][using {\scriptsize THINGS} rotation curves from \citealt{DeBlok2008} to estimate $H$]{Bagetakos2011} in local star-forming galaxies, and their predicted outcomes.} Orange and hatched grey regions denote parameter space where the bubbles \rev{are expected to} have broken out of the ISM disk ($R_b > H$) or fragmented within it ($v_b < \sigma$), respectively.  Bubble radii and velocities (Eqs. 7 \& 8 O21, see also \S~\ref{sec:remnantevolution}) \rev{are predicted to evolve in $\log v_b/\sigma$-$\log R_b/H$ along lines of constant slope ($-(2+\alpha)/(2-\alpha)$ in log-log space). The dashed line in the `evolving bubbles' portion of parameter space demarcates the superbubbles we predict will fragment within the ISM (\textbf{PS case}) from those that this model predicts will break out of the local gas disk (\textbf{PBO case}).  \textbf{Bolded points} indicate superbubbles in regions with inverse dynamical times $\Omega > 50$ Gyr$^{-1}$ (or $t_{\rm dyn} < 20$ Myr $= t_{\rm SNe}/2$), showing that almost all of the `largest' bubbles (relative to local $H$) have short dynamical times (\ie in the central regions of galaxies).}}
    \label{fig:rbvb}
\end{figure}

 \rev{In Figure~\ref{fig:rbvb}, we plot observed HI bubble radii and expansion velocities from local star-forming galaxies in the {\scriptsize THINGS} survey by \citet[][we used {\scriptsize THINGS} rotation curves from \citealt{DeBlok2008} to estimate local $H$]{Bagetakos2011} to interpret the likely outcome of these bubbles. \citet{Bagetakos2011} identified gaps and voids in spatially and velocity-resolved data of the {\scriptsize THINGS} HI disks, fitting ellipsoids to find HI bubble sizes and expansion velocities. Their sample was divided up into three types of HI bubble, here we consider only their \emph{`type 3'}, where the bubble is still intact with both near and far edges detected, comparable with still-evolving superbubbles in our model.}

\rev{Predicting whether or not we expect an observed superbubble to fragment or break-out is possible when considering the $t<t_{\rm SNe}$ cases of Eq.~7 \& 8 of O21 (see \S~\ref{sec:remnantevolution}).  Taking the ratio $v_b/R_b$, we find that this evolves as $v_b \propto R_b^{-(2+\alpha)/(2-\alpha)}$.  And so, bubbles observed to be below a line of this constant slope divide the $v_b$-$R_b$ space into bubbles that we expect to fragment and those that we expect to break-out.}

\rev{The observed bubbles whose radii appear to already be greater than $H$ all come from short dynamical time regions, where $1/\Omega = t_{\rm dyn} < 20$ Myr $ = t_{\rm SNe}/2$. This suggests that the HI bubbles are remaining fairly coherent after they have already broken-out of the galactic nuclei, or that we are systematically underestimating $H$ in these regions.  A number of these bubbles are in the inner ring of NGC 4736, which appears to be a dynamically induced starburst \citep{Munoz-Tunon2004}, for which our assumption of $\tilde Q_{\rm gas} \approx 1$ may not hold. As well, identifying \emph{intact} HI bubbles here may be problematic given the predominantly molecular nature of the central regions of local $L_\star$ star-forming galaxies \citep{Jimenez-Donaire2019}. }

\rev{Considering the observed intact bubbles from regions with $\Omega < 50$ Gyr$^{-1}$ (for $v_c \approx 200$ km/s, this is $R > v_c/\Omega \approx 4$ kpc), the ensemble of radii and expansion velocities suggests that all of these HI bubbles will eventually fragment in the ISM rather than drive significant outflows/fountains.  This is consistent with the $\tilde f_{\rm gas}$-$\Omega$ profiles of local star-forming galaxies (see Figure~\ref{fig:case}), where the only regions that host superbubble breakout are the central starbursts (where various gas dynamics have fed the formation of central super star clusters).}

\section{Discussion \& Summary}\label{sec:discussion}

\subsection{The Extreme Rarity of Coasting Outcomes}\label{sec:critpt}

As discussed in \S~3.4 of O21, the primary difficulty in realizing coasting (\textbf{CBO} and \textbf{CF cases}) outcomes appears to lie in the fact that the star-forming extent of the vast majority of galaxies does not reach so far out (radially) to have dynamical times exceeding a 100 Myr (\ie $1/t_{\rm dyn} = \Omega = 10$ Gyr$^{-1}$).  For the most part, rotation curves in galaxies are sufficiently high, and their star-forming edges sufficiently close, that dynamical times remain shorter than $4 t_{\rm SNe}/(2-\alpha)$, and only \emph{powered} outcomes are seen.  The only exceptions may be in ultra-diffuse dwarfs (UDGs), having low $v_c$ and large extents \citep[][]{Beasley2016}.  Even then, this model is only relevant for those which still maintain some star-forming gas (improbable for UDGs).

\subsection{Alternative Cluster Formation Model: Constant Star Cluster Formation Efficiency}\label{sec:constanteff}
\begin{figure}
	\includegraphics[width=0.97\columnwidth]{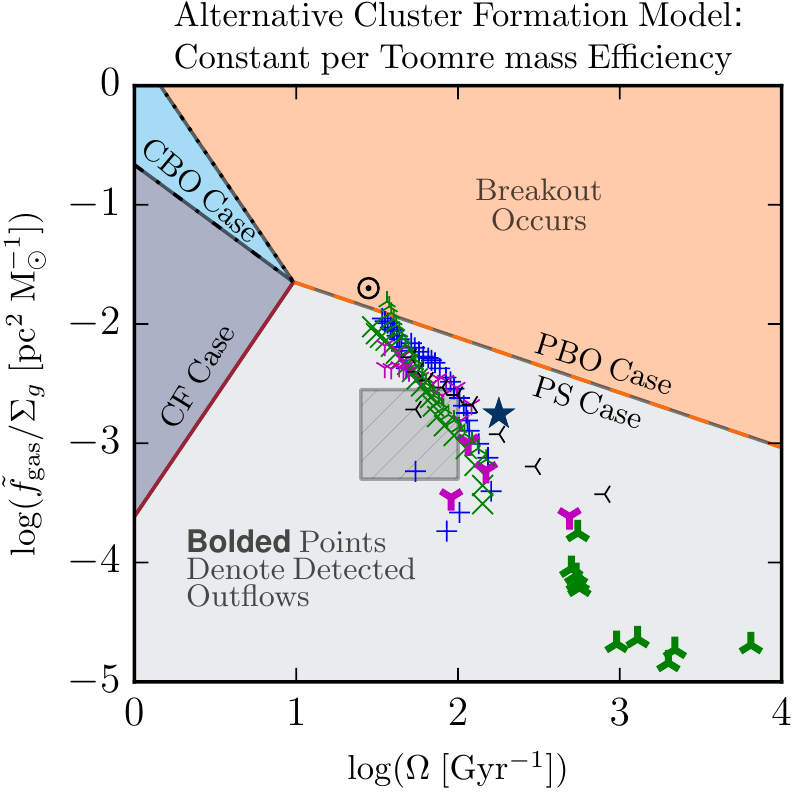}
    \caption{Case boundaries with constant star cluster formation efficiency (per Toomre-mass) assumed for the superbubble model, in the style of Figure~\ref{fig:case}, in inverse disk surface density--dynamical time phase space, following required adjustments to Eqs.~9, 12, 13 \& 15.  In replacing $\Sigma_{\rm crit} \rightarrow \Sigma_g/\epsilon_{\rm cl}$, the adjustment to the case boundaries is found by substituting $\tilde f_g \rightarrow \tilde f_g/\Sigma_g$ (\emph{i.e.}, $1/\Sigma_{\rm disk}$). With a constant star cluster formation efficiency, \textbf{CBO/CF cases} remain unlikely to occur given the long dynamical times required (unchanged from fiducial model predictions).  In this case, our fiducial model parameters and $\epsilon_{\rm cl}=0.01$ predict that breakout occurs at lower disk surface densities (almost always $<$130 \mpcsqt), implying that breakout,  
    \ie outflows/fountains never occur in the inner disk, contrary to observations. In spite of the outflow detection in NGC 4321 by \citet{Castillo-Morales2007}, and the well-known superwind of NGC 253, we would not predict the centers of either NGC 253 or 4321 to host a SN-driven outflow. Instead, this constant efficiency model predicts outflows from Solar Circle (and more diffuse) disk conditions.  Moreover, we would also expect the surface densities in high-redshift SMGs to be such that there are no predicted outflows, contrary to observations from $z \sim 1-3$.}
    \label{fig:cases_constanteff}
\end{figure}
Alternative models for star cluster formation efficiency have been proposed, arguing that star formation proceeds at a constant efficiency of roughly 1\% per free-fall time (\citealt{Krumholz2007a}; see \citealt{Krumholz2018c} for a review of (low) star formation efficiency in clusters).  If we were to adopt a constant-efficiency per Toomre-mass model, as opposed to the \citet{Grudic2018} model, where $M_{\rm cl} = \epsilon_{\rm cl} M_g \approx \epsilon_{\rm cl} \pi H^2 \Sigma_g$, where $\epsilon_{\rm cl} = 0.01$, and holding the rest of the model fixed, then throughout the text the only difference required would be to replace $\Sigma_{\rm crit} \rightarrow \Sigma_g/\epsilon_{\rm cl}$.  Whereupon, we would plot all our figures in $\tilde f_g/\Sigma_g$--$\Omega$ space, rather than $\tilde f_g$--$\Omega$ space.  The rest of the results and analysis would remain unchanged. Figure~\ref{fig:cases_constanteff} shows the case boundaries in the somewhat unusual $\tilde f_g/\Sigma_g$--$\Omega$ space.

The main difference of this alternative model is that breakout is now dependent on the local \emph{disk} surface density, as $\Sigma_{\rm disk} \approx \Sigma_\star + \Sigma_g = \Sigma_g/\tilde f_g$.  In fact, such a model would imply that breakout only occurs at lower (relatively speaking) disk surface densities for a given local dynamical time: above a local disk surface density, the resulting superbubble is effectively smothered by the disk.  At the critical dynamical time (where all four case boundaries intersect) this local disk surface density is $\approx$45 \mpcsqt, and it grows to $\approx$130 \mpcsqt~at~$\Omega = 10^2$ Gyr$^{-1}$.  Given that regions with shorter dynamical times are generally core-ward in disk galaxies, and that generally the disk (stellar + gaseous) surface densities of the central regions in Milky Way mass galaxies greatly exceed 100 \mpcsqt~(see the PHANGS-ALMA sample of \citealt{Sun2020}), this model suggests that superbubble breakout, and thus outflows, would not occur in the inner disk regions of galaxies, only in their outskirts.  This is contrary to many observations of galactic winds and outflows, specifically those of galaxies without AGN, which nonetheless tend to report outflows and fountains originating from the inner regions of galaxies \citep[\eg][]{Bolatto2013a}.  Consequently, these data disfavor a constant (per Toomre-mass) star cluster formation efficiency within this superbubble feedback model.

\subsection{Summary}\label{sec:conclusion}

In this Letter, we compared a model (derived in O21) of clustered SNe feedback in disk environments with observations of local and high-redshift star-forming galaxies.  Of specific interest, we tested our predictions from O21 of the ability of supernova-driven superbubbles to break out of the gas disk of a galaxy with known hosts of superwinds, and galaxies thought to lack them. As well, we examined observed HI bubbles/holes in the context of our predicted scalings for bubble radii and velocities.

Key takeaways from comparing this model to observations include:
\begin{itemize}

\item Spatially resolved observations of $z\approx 0$ star-forming galaxies suggest that most star-forming regions in the local universe fall into the `\textbf{PS case}', \ie that superbubbles stall and fragment inside the disk and locally deposit almost all of their momentum.  Higher-redshift observations suggest that $z \sim 2$ SMGs exist in `\textbf{PBO case}' parameter space, \ie superbubbles at $z\sim 2$ are (always) able to drive outflows/fountains.  The central regions of some local galaxies also appear to lie in the predicted `\textbf{PBO case}' region (\eg NGC 4321, which has evidence of central star formation-driven winds).  The transition from high to low redshift galaxies, in terms of hosting pervasive outflows to only those in circumnuclear regions, appears driven by an evolution from high to low local gas fractions in star-forming regions according to this model.

\item \rev{Observed intact HI bubble radii and velocities in local star-forming galaxies (from \citealt{Bagetakos2011}) are consistent with the $\tilde f_{\rm gas}$-$\Omega$ profile interpretations: most feedback driven bubbles in local galaxies should fragment inside the ISM, and that those able to break-out originate from short dynamical time regions in the nuclear regions.}

\item A cluster formation model that includes a constant star formation efficiency per Toomre mass is effectively ruled out by the observational data (see \S~\ref{sec:constanteff}), as this model predicts that high surface density regions (\eg high-redshift star-forming clumps or low-redshift circumnuclear regions) would be unable to host superbubbles capable of breaking out and driving outflows/fountains.

\end{itemize} 

In comparing to observations, we find that the clustering of supernovae indeed has \rev{important} implications for the local efficacy of star formation, and the evolution of galaxies more broadly across cosmic time.  Future highly spatially resolved observations, capable of identifying and quantifying the properties of supernova-driven superbubbles\rev{, especially in dense molecular gas structures,} should help to further constrain the effective strength of feedback under varying local galactic conditions and inform sub-grid models for feedback in cosmological galaxy simulations.


\begin{acknowledgments} 
MEO is grateful for the encouragement of his late father, SRO, in studying astrophysics.
We thank Alex Gurvich, Lee Armus, and Phil Hopkins for conversations relating this model to disk formation at intermediate redshifts, and connections with spatially resolved observations and superwinds.
\rev{We also thank the anonymous referee for comments and suggestions that significantly strengthened the manuscript.}
MEO was supported by the National Science Foundation Graduate Research Fellowship under Grant No. 1144469.
The Flatiron Institute is supported by the Simons Foundation.
This research has made use of NASA's Astrophysics Data System.
B.B is grateful for support from the Packard Fellowship and Sloan Fellowship.
\end{acknowledgments}




\bibliographystyle{aasjournal} 
\bibliography{library, alt_bibs} 




\label{lastpage}
\end{document}